\documentclass[a4paper,11pt]{article}
\pdfoutput=1
\usepackage{jcappub}

\usepackage[T1]{fontenc}

\usepackage{multirow}

\title{\boldmath Strong lensing as a probe of braneworld}

\author[a,1]{Yi  Zhang,\note{Corresponding author.}}
\author[a]{Hong Liu,}
\author[a]{Dan Wen,}
\author[b]{Hongsheng Zhang }

\affiliation[a]{ College of  Science, Chongqing University
of Posts and Telecommunications,\\ Chongqing 400065, China}
\affiliation[b]{School of Physics and Technology, University of Jinan, 336, West Road of Nan Xinzhuang, \\Jinan 250022, Shandong, China}

\emailAdd{zhangyia@cqupt.edu.cn}

\abstract{ For the first time, we use the Event Horizon Telescope (EHT) data to  constrain  the  parameters  of braneworld black holes which constrain   $\epsilon=0.0285^{+0.0888+0.1456}_{-0.0895-0.1475}$ for the anisotropic black hole and  $q=-0.0305^{+0.1034+0.1953}_{-0.0895-0.1470}$  for the tidal Reissner-Nordstr$\ddot{o}$m (RN) black hole.    Based on  the fitted data and physical requirement, we calculate the photon deflection, the angular separation and time delay between different relativistic images of the  anisotropic black hole  and the tidal RN black hole in  the  ranges $-0.1190<\epsilon<0$ and  $-0.1775 <q<0$.   And furthermore,  we study the  quasinormal modes (QNMs) for the braneworld black holes.  The results  shed light on existence of extra dimension.    }

\begin{document}
\maketitle
\flushbottom

\section{Introduction}
 Black  hole  is one of the most exciting predictions of Einstein's general relativity  whose  existence  is    confirmed   directly by   capturing images of the black hole shadow  from the  Event horizon Telescope (EHT) \cite{EHT1,EHT2,EHT3,EHT4}.
 Theoretically, it is widely accepted that general relativity is an effective infrared gravitational theory and should be modified in the ultraviolet regime \cite{Will:2005va,Jordan:2008ky,Turyshev:2007qy,Turyshev:2008dr}.
To extend the theory to high energy regime,  a natural way is to introduce extra  dimension, which  may play a significant role in  unification theory and quantum gravity.        Braneworld is one of the most popular models for extra dimension \cite{Randall:1999vf}.  The braneworld paradigm views our universe as a slice of some higher dimensional spacetime. Unlike the Kaluza-Klein picture of  extra dimensions,  where we do not  sense the  extra dimensions because they are  so small and our  physics is ``averaged'' over them. The braneworld   picture can have large, even non-compact but highly warped extra dimensions which are unobservable at low energy region since the gauge fields are confined to the brane.
This scenario provides a set-up in  which  we have standard four dimensional physics confined to the brane, while gravity can propagate in  the bulk. Black holes within the braneworld framework may exhibit significant potential differences from those in general relativity \cite{cham}.

Technically, by using  Gauss-Codazzi approach  \cite{Shiromizu:1999wj}, the classical five dimensional braneworld black hole solution is reduced  to the  four dimensional quantum radiating black hole.  But  the exact  metric describing the spacetime geometry around braneworld black holes is not yet  known. Here, we concentrate on
the anisotropic  and     tidal Reissner-Nordstr$\ddot{o}$m      black holes. The anisotropic black hole  describes the properties near the horizon \cite{Gregory:2004vt,Tanaka:2002rb,Creek:2006je,Maartens:2000fg}. Since this  system contains  unknown bulk dependent term, assumptions have to be made either in the form of metric by the Weyl  term.
This braneworld black hole  is believed to encode  quantum correction of black holes.    Another workable solution in braneworld is given in \cite{Dadhich:2000am} , which is called     tidal Reissner-Nordstr$\ddot{o}$m      black hole. The reason for neglecting the Garriga-Tanaka metric is that its validity is confined to the far-field limit\cite{Garriga:1999yh,Giddings:2000mu}.

The highly bending, even looping of light rays around black holes in strong fields is a well-known and amazing predictions of general relativity \cite{Dyson:1920cwa,Zwicky:1937zzb,Einstein,Schneider,Petters,Perlick,Schneider2006,Darwin,Virblack holeadra:1999nm,Bozza:2001xd,Bozza:2002zj} .   It is significant and interesting to investigate the braneworld   effects through observations of EHT and thus presents effective constraints on braneworld models.
 In strong field,  the light  deflection divergences at photon sphere. By an analytic approximation method,  Bozza proved  that, when the angle between  source and  lens tends to zero, the deflection angle diverges logarithmically. Bozza et al. \cite{Bozza:2001xd,Bozza:2002zj}  find an interesting simplification for the lens equation in such regime, finding the expression for observable quantities in the so-called strong deflection limit  regime. And theoretically, light bending by a compact body  can exceed $2\pi$ and the light even can wind several loops before escaping, which develops infinite discrete images on two sides of the body closely, called relativistic images. See  Refs.\cite{Bozza:2010xqn,AbhishekChowdhuri:2023ekr} for more details.  Relativistic images, which are not predicted by the classical weak gravitational lensing, provide a new way to study the properties of spacetime in the strong gravitational field.  Differences in the deflection angle are significantly reflected on the relativistic images.   In recent  years,  more approaches have been developed, such as   time delay  \cite{Bozza:2003cp,Cavalcanti:2016mbe,Ghosh:2020spb}  and  QNM \cite{Cardoso:2008bp,Dolan:2009nk,Stefanov:2010xz,Bohra:2023vls}. The various strong deflection lensing work  can been seen in Refs.\cite{huetal, morehu,Tsukamoto:2016jzh,Zhao:2016kft,Afrin:2021wlj,Majumdar:2004mz,Nandi:2006ds,Tsukamoto:2016qro,deXivry:2009ci,Gralla:2019xty,Ghosh:2022mka,Soares:2023err,Soares:2023uup,Hsieh:2021scb,Hsieh:2021rru}.
    Based on the property that the divergence of deflection angle can be integrated up to first order, one derives the gravitational lensing observables.  Using the  M87* and Sgr A*   black hole shadow data, one can investigate the parameters  of braneworld black holes. The $\chi^2$-test is an effective  method which extracts information from the observational data to obtain the black hole parameter range   \cite{Kumar:2018ple,Pantig:2022ely,Afrin:2021wlj}.

 The paper is structured as follows: In section \ref{black hole},  we introduce the two braneworld black holes:  the anisotropic  and the tidal Reissner-Nordstr$\ddot{o}$m      ones. Then we apply the strong field limit   procedure \cite{Bozza:2002zj} to the braneworld metrics  in sections \ref{gl} and \ref{Bozza}.
  In section \ref{obs}, we calculate the observation effects,  including the positional separations $\theta_{\infty}$, brightness difference  $s$ and magnitude difference  $r$.  Furthermore,   we will discuss quasi normal mode  and the time delay of both images as well.
  At last, we give a   conclusion in section \ref{conclusion}.

\section{The    metric on  the brane}\label{black hole}
Randall and Sundrum showed that a four dimensional Minkowskian braneworld can be constructed  although gravity was inherently five dimensional, where the spacetime was strongly warped.
In the Randall-Sundrum II model   a  single membrane of  positive tension imbedded in five dimensional AdS  space,
\begin{eqnarray}
ds^2=g_{ab}dx^a dx^b =d\tilde{z}^2+a^2(\tilde{z})\eta_{\mu\nu}dx^{\mu}dx^{\nu}.
\end{eqnarray}
Here,  $a(\tilde{z})=e^{-|\tilde{z}|l}$, where $l$  is the radius of the AdS, and $\eta_{\mu\nu}$ is the Minkowski metric in four dimension.
    The Randall-Sandrum II model offers a remarkable compactification, that is, on scales
much larger than $l$, four dimensional gravity is recovered on the brane. For some five dimensional braneworld solutions, the difference in the observables is found to be rather small from the four dimensional  Schwarzschild  case \cite{Whisker:2008kk,Pantig:2022ely,Dahia:2014tea,Whisker:2004gq,Keeton:2006di,Bin-Nun:2009hct,Eiroa:2004gh}.

Considering a static metric,   there exists  a  five dimensional solution  analogous to  the C-metric  in four dimensions which  has  a timelike Killing vector,  and  can therefore be ``sliced''  by the  braneworld in  such  a way  as to create  a static four dimensional black hole on the brane \cite{Gregory:2004vt,Tanaka:2002rb,Creek:2006je,Maartens:2000fg}.
Here, the spacetime is constructed  so that there are four dimensional flat slices stacked along the fifth $\tilde{z}$-dimension, which have a $\tilde{z}$-dependent  conformal pre-factor known as the warp  factor. This  warp factor has a cusp at  $\tilde{z}=0$, which indicates the presence of a domain wall, or the braneworld, which represents an exact flat Minkowski universe \cite{Gregory:2004vt}.
  The Einstein  equation for the simplest case (``vacuum'' brane)  can be written as,
  \begin{eqnarray}
  \label{einsteineq}
 G_{\mu\nu}={\cal E}_{\mu\nu}.
 \end{eqnarray}
where the   ${\cal E}_{\mu\nu}$   is the Weyl  term, consisting of projection of the bulk Weyl tensor on  the brane.
 In the  AdS picture,  the brane is not at the  AdS boundary,  but  at  a finite  distance, and the theory on the brane now contains a conformal energy-momentum tensor, which appears as the Weyl term ${\cal E}_{\mu\nu}$. Using the symmetry of the physical set-up to put the Weyl energy  into  the form   \cite{Gregory:2004vt,Tanaka:2002rb,Creek:2006je,Maartens:2000fg},
\begin{eqnarray}
{\cal E}_{\mu\nu}={\cal U}(u_{\mu}u_{\nu}-\frac{1}{3}h_{\mu\nu})+\Pi(r_{\mu}r_{\nu}+\frac{1}{3}h_{\mu\nu}),
\end{eqnarray}
where $u_{\mu}$ is a unit time vector,  $r_{\mu}$ is a unit radial vector, and $h_{\mu\nu}$  is the metric perturbation.
Then    the field equations are
\begin{eqnarray}
&&G_{t}^{t}={\cal U},\\
&&G_{r}^{r}=  -\frac{{\cal U}+2\Pi}{3},\\
&&G_{\theta}^{\theta}= -\frac{{\cal U}-\Pi}{3},
\end{eqnarray}
Where  ${\cal U}$ and $\Pi$ are the Weyl energy and the anisotropic stress, respectively.

 \subsection{The anisotropic  metric}
The simplest solution  of  Eq.(\ref{einsteineq}) is  based on   the static spherically  symmetric  metric on the brane  which is
\begin{eqnarray}
ds^2=-A(r)dt^2+B(r)dr^2+C(r)d\Omega_{\uppercase\expandafter{\romannumeral2}}^2
\end{eqnarray}
where $C(r)=r^2$ and $d\Omega_{II}^2= d\tilde{\theta} ^2+\sin^2\tilde{\theta}  d\tilde{\phi}^2$.
And, the horizon is the asymptotic regime in which we could withdraw some information  about the black hole.
Then,  when ${\cal U}=0$,  by  setting  $G=1$,  there is a simple analytic solution which is   near horizon   in  area gauge \cite{Gregory:2004vt} ,
\begin{eqnarray}
\label{area}
ds^{2} =-[(1+\epsilon)\sqrt{1-\frac{2M}{R}}-\epsilon]^2dt^2+(1-\frac{2M}{R})^{-1}dR^2+R^2d\Omega_{\uppercase\expandafter{\romannumeral2}}^2,
\end{eqnarray}
where $R=(r+r_0)^2/r$, $M=2r_0$, $\epsilon =(-r_1+r_0)/M$ and $r_1$ is the integral constant.  And  based on \cite{Gregory:2004vt}, it also  has another form as
 \begin{eqnarray}
 \label{iso}
ds^{2} =-\frac{(r-r_1)^2}{(r+r_0)^2}dt^2+\frac{(r+r_0)^4}{r^4}dr^2+\frac{(r+r_0)^4}{r^2}d\Omega_{\uppercase\expandafter{\romannumeral2}}^2.
\end{eqnarray}
 This metric describes the behaviors   in  the near   horizon regime. When the metric ${\cal U}=0$,  the  anisotropic stress  for this solution is $\Pi=3M\epsilon/AC^3$.
 For convenience,  we call such a metric as the   anisotropic  metric which is based on  the non-perturbative nature of gravity.  When $\epsilon=0$ which  corresponds to ${\cal E}_{\mu\nu}=0$, it is  back to  Schwarzschild  BH (the $R=2GM$,and $G=1$).
 And, the area gives a familiar spatial part  of  the metric, for $\epsilon>0$, $g_{tt}$ will be zero before $r=2M$, and the  area  gauge holds outside the  black hole. Then, this solution could not be treated as black hole, because the   `horizon' is singularity.
Here, we take $M$ as the measure of distances. After defining $x=R/2M$, the anisotropic metric in area gauge which shows near horizon modification  to general relativity is written as follows,
\begin{eqnarray}
ds^{2} =-[(1+\epsilon)\sqrt{1-\frac{1}{x}}-\epsilon]^2dt^2+(1-\frac{1}{x})^{-1}dx^2+x^2d\Omega_{\uppercase\expandafter{\romannumeral2}}^2.
\end{eqnarray}

\subsection{The tidal Reissner-Nordstr$\ddot{o}$m   metric}

As  the Weyl term ${\cal E}_{\mu\nu}$ is antisymmetric  and    trace-free  which behaves as the stress-energy $T^{em}_{\mu\nu}$ of a Maxwell field,    the  tidal Reissner-Nordstr$\ddot{o}$m  could be \cite{Dadhich:2000am},
\begin{eqnarray}
ds^{2} =-(1-\frac{2M}{r}+\frac{Q}{r^2})dt^2+(1-\frac{2M}{r}+\frac{Q}{r^2})^{-1}dr^2+r^2d\Omega_{\uppercase\expandafter{\romannumeral2}}^2.
\end{eqnarray}

 Rescaling the radius coordinate by the mass $x=r/2M$, the tidal Reissner-Nordstr$\ddot{o}$m  metric is written as,
\begin{eqnarray}
	ds^{2} =-(1-\frac{1}{x}+\frac{q}{x^2})dt^2+(1+\frac{1}{x}+\frac{q}{x^2})dx^2+x^2d\Omega_{\uppercase\expandafter{\romannumeral2}}^2.
   \label{RN}
	\end{eqnarray}


\section{A  general introduction to strong lensing limit approach }\label{gl}

In this section, for convenience  we give a brief review on the general formula of gravitational lensing in the strong field limit.    Due to the spherical symmetry, we only consider light rays moving on the equatorial plane with $\tilde{\theta}=\frac{\pi}{2}$.
 The lens equation is used to define the geometrical relations among the observer, lens and  source,
which generally can be written as  \cite{Bozza:2001xd}
\begin{eqnarray}
\tan \beta= \tan \theta-\frac{D_{LS}} {D_{OS}}[\tan \theta +\tan (\theta-\alpha)]
\end{eqnarray}
where $\alpha$ is the deflection angle, and $\beta$ is the angular separation between the source and the lens, $\theta$ is the angular separation between the image and the lens, and $D_{LS}$ and $D_{OS}$ are  the projected distance of lens-source and observer-source along the optical axis. Given  a source position $\beta$, by solving this equation, the value of $\beta$  denotes the  position  of the images observed by O.

We assume that both the observer and the source are far from the lens and  the spacetime of the lens is asymptotically flat.     We shall pay attention to situations where the source is almost perfectly aligned with the lens. In this case,  we are allowed to expand $\tan \beta$ and $\tan \theta$ to the first  order. With $\alpha= 2n\pi + \Delta \alpha_n$, and $n$ integer, we can perform the expansion $\tan (\alpha-\theta)$ $\sim$ $\Delta \alpha_n-\theta$.
	After assuming $\alpha,\beta,\theta\ll 1$, the lens equation becomes
\begin{equation}
\label{smallangle}
\beta=\theta-\frac{D_{LS}}{D_{OS}}\Delta \alpha_n=\theta-\frac{D_{LS}}{D_{OS}}(\alpha(\theta)-2n\pi),
\end{equation}
where $\Delta \alpha_n=\alpha(\theta)-2n\pi$   is the extra angular deflection angle after a photon with a deflection angle winding $n$ loops.
The deflection angle $\alpha$  encodes the physical  information about the deflector which can be calculated  through the integration of the geodesic  of the light ray.
Due to the  asymptotically approximated lens equation, the spacetime of the lens only  affects the  deflection angle $\alpha(\theta)$ which will be calculated in the strong lensing.

Conserved quantities along the orbit are $E=A(r)\dot{t}$ and $L=C(r)\dot{\tilde{\phi}}$, where a dot denotes derivative with respect to the affine parameter. Considering the conservation of energy and angular momentum
\begin{eqnarray}
\frac{d\phi}{dx}= \frac{\sqrt{B}}{\sqrt{C}\sqrt{\frac{CA_0}{C_0A}-1}},
\end{eqnarray}
we find
\begin{eqnarray}
I(x_0)=\int^\infty_{x_0} \frac{\sqrt{B}}{\sqrt{C}\sqrt{\frac{CA_0}{C_0A}-1}} dx,
\end{eqnarray}
where $x_0$ is the closest distance of the photon to the black hole, and $A_0$ and $C_0$ are the values of $A$ and $C$ when $x=x_0$.
The deflection angle for the null geodesic of a photon in the black hole spacetime can be found
\begin{eqnarray}
\alpha(x_0)=-\pi +\int_{x_0}^{\infty} \frac{2\sqrt{B}dx}{\sqrt{C}\sqrt{\frac{CA_0}{C_0A}-1}}.
\end{eqnarray}
When the light ray trajectory gets closer to the event horizon, the deflection angle increases.

   \section{The Bozza's procedure}\label{Bozza}

We follow the Bozza's procedure\cite{Bozza:2002zj} to discuss the strong lensing problem.
It   has  been proved that when a photon moves around a black hole, there exists an innermost unstable orbit named as photon sphere.
First, we calculate $x_m$, which  is the largest  root of the following equation
\begin{equation}
\frac{C'(x)}{C(x)}=\frac{ A'(x)}{A(x)},
\label{rm}
\end{equation}
where $A$,  $C$, $A'$ and $C'$ must be positive for $x>x_m$.
And, this equation admits at least one positive solution. We shall call $x_m$
 the radius of the photon sphere.    The deflection angle is divergent at  the photon sphere $x_m$.

We introduce the impact parameter $u=L/E$ which is the perpendicular distance from the center of the mass of  lens to the tangent of the null geodesics and remains constant throughout the trajectory. By conservation of the angular momentum, the closest distance is related to the impact parameter by
\begin{eqnarray}
\label{u}
u=\sqrt{\frac{C_0}{A_0}}=|\frac{L}{E}|,
\end{eqnarray}
where the subscript $0$ indicates that the function is evaluated at $x_0$.
To expand the integral near the photon sphere not only provides an analytic re-presentation of the deflection angle but also  shows the behavior of photons near the photon sphere.
 The  minimum value of $u$  could be  written as
 \begin{eqnarray}
 u_m=\sqrt{\frac{C_m}{A_m}},
 \end{eqnarray}
 where $A_m$ and $C_m$ are the values of $A$ and $C$ when $x=x_m$.
 The track of a photon incoming from infinity with some impact parameter $u$ will be curved while approaching the black hole.   And $\alpha$  higher than $2\pi$ will result in loops of the light ray around the black hole.

Then, we define $y=A(x)$  and
$z=(y-y_{0})/(1-y_{0})$
and rewrite the integral $I(x_{0})$ as
\begin{eqnarray}
\label{I(x_{0})}
&&I(x_{0})=\int_{0}^{1}R(z,x_{0})f(z,x_{0})dz
\end{eqnarray}
 where
 \begin{eqnarray}
&&R(z,x_{0})=\frac{2\sqrt{By}}{CA^{\prime }}\left( 1-y_{0}\right) \sqrt{C_{0}},\\
&&f(z,x_{0})=\frac{1}{\sqrt{y_{0}-[(1-y_{0})z+y_{0}]\frac{C_{0}}{C}}}.
\end{eqnarray}
The function $R(z,x_{0})$ is regular for all
values of its arguments, but the function $f(z,x_{0})$ diverges as $
z\rightarrow 0$.  Following  the Bozza's procedure\cite{Bozza:2002zj} ,   the argument of the square root in $f(z,x_0)$ is expanded to the second order in $z$, and then
\begin{eqnarray}
 f (z,x_0) \sim f_0(z,x_0)= \frac{1}{\sqrt{\tilde{\alpha} z +\tilde{\beta} z^2}}
\end{eqnarray}
where
\begin{eqnarray}
&& \tilde{\alpha}  =\frac{1-y_0}{C_0 A'_0}  \left(C'_0 y_0-C_0 A'_0,
\right) \\
&& \tilde{\beta}  = \frac{\left( 1-y_0 \right)^2}{2C_0^2 {A'_0}^3} \left[
2C_0 C'_0 {A'_0}^2 + \left(C_0 C''_0-2{C'_0}^2 \right)y_0 A'_0 -C_0 C'_0 y_0
A''_0 \right].
\end{eqnarray}
To get the integral of the Eq.(\ref{I(x_{0})}), we divide the divergence into a regular part and a divergent one, which could be
\begin{eqnarray}
I(x_0)=I_D(x_0)+I_R(x_0) ,
\end{eqnarray}
where $I_D$ and $I_R$ are the divergent part and the regular part,respectively, $x_0$ is the closet distance, and
\begin{eqnarray}
\label{ID}
&&I_{D}(x_0)=\int_{0}^{1}R(0,x_m)f_0(z,x_0)dz,\\
\label{IR}
&&I_{R}(x_0)=\int_{0}^{1}g(z,x_0)dz=\int_{0}^{1}(R(z,x_{m})f(z,x_{m})-R(0,x_{m})f_{0}(z,x_{m}))dz,
\end{eqnarray}
 where the latter gives the deflection angle to order $O(x_{0}-x_{m})$,   the function $g(z,x_{m})$ is regular at $z=0$,     and    the $f_0(z, x_0)$ is the expansion of the parameter of the square root in $f(z, x_0)$ to the second order at $z$.

At last, we expand   $u$ defined in  Eq.(\ref{u}),
\begin{eqnarray}
\label{um}
u-u_m=\hat{c}(x-x_m)^2
\end{eqnarray}
where $\hat{c}=(C''_my_m-C_mA''_m)/(4\sqrt{y_m^3C_m}) $.
Then, by using $u_m$ instead of $x_m$,  the deflection angle could be expressed as
\begin{eqnarray}
\label{alphaeq}
\alpha (\theta)=-\bar{a}\ln \left( \frac{\theta D_{OL}}{u_{m}}-1\right) +\bar{b},
\end{eqnarray}
where
\begin{eqnarray}
\label{a}
&&\bar{a}=\frac{R(0,x_{m})}{2\sqrt{\beta _{m}}},\\
\label{b}
&&\bar{b}=b_0+b_{R}=-\pi +\bar{a}\ln \frac{2\beta _{m}}{y_{m}}+b_{R},
\end{eqnarray}
where   $b_R=I_R(x_m)$ and $\beta _{m}=2\hat{c}C_ m^{3/2}y_m^{-1/2}(1-y_m)^2$.    $\bar{a}$ and $b_0$  encode the divergent  part, and $b_R$ represents the regular part.
Then, the two integrals (Eqs. (\ref{ID}) and (\ref{IR}))
are expanded around the photon sphere $x_m$.
As  the value of $x_m$ is known, we could compute $\alpha$ by  a proper  expansion in  the parameters of the metric.

\section{The observables}\label{obs}
 In this section, we will show how to translate the parameters $u_m$, $\bar{a}$ and $\bar{b}$   to the observables. Then, through  the observations of strong lens, we obtain the metric parameters ,  which could probe the space time structure.  We consider  the simplest condition   where outermost image $\theta_1$ is resolved as a single image, while all the remaining ones are packed together at $\theta_{\infty}$.
 The strong field gravitational lensings  are helpful to distinguish different types of black holes if we can separate the outermost relativistic images and determine their angular separation, brightness difference, time delay and QNM.

\subsection{ The parameter estimation from the positional separation  $\theta_{\infty}$ }\label{constrained}
Theoretically, when the lens and  observer are nearly aligned and the black hole has spherical symmetry,  we can define   the angular radius of shadow of black hole  as
\begin{eqnarray}
\theta_{\infty}=\frac{u_m}{D_{OL}}.
   \end{eqnarray}
     Inversely,  the impact parameter could be detected by  the angular radius of shadow of black hole with
$u_{m}=\theta_{\infty}D_{OL}$.  In observation, for M87*,
   the shadow angular diameter is $ \theta_{\infty} = 21\pm 1.5 \mu as$, the distance of the M87* from the Earth is $D_{OL}= 16.8 Mpc$, and the mass of the M87* is $ 6.5 \pm 0.90\times10^9 M_{\odot}$. For Sgr. A* the shadow angular radius is  $\theta_{\infty} = 24.35 \pm 3.5 \mu as $(EHT), the distance of the Sgr. A*  from the Earth is $D_{OL} = 8277\pm33 pc$ and the mass of the black hole is $ 4.3 \pm0.013\times10^6 M_{\odot} $(VLTI).
Then, to discuss the observational constraint on $\theta_{\infty}$ by using the  data from M87* and Sgr A* of EHT, we make $\chi^2$ test which is defined as
\begin{eqnarray}
\chi^2=\frac{(\theta^{theory}_{\infty}-\theta^{observation}_{\infty})^2}{error^2}.
\end{eqnarray}

\begin{figure}[ht]
\centering
\includegraphics[width=5cm]{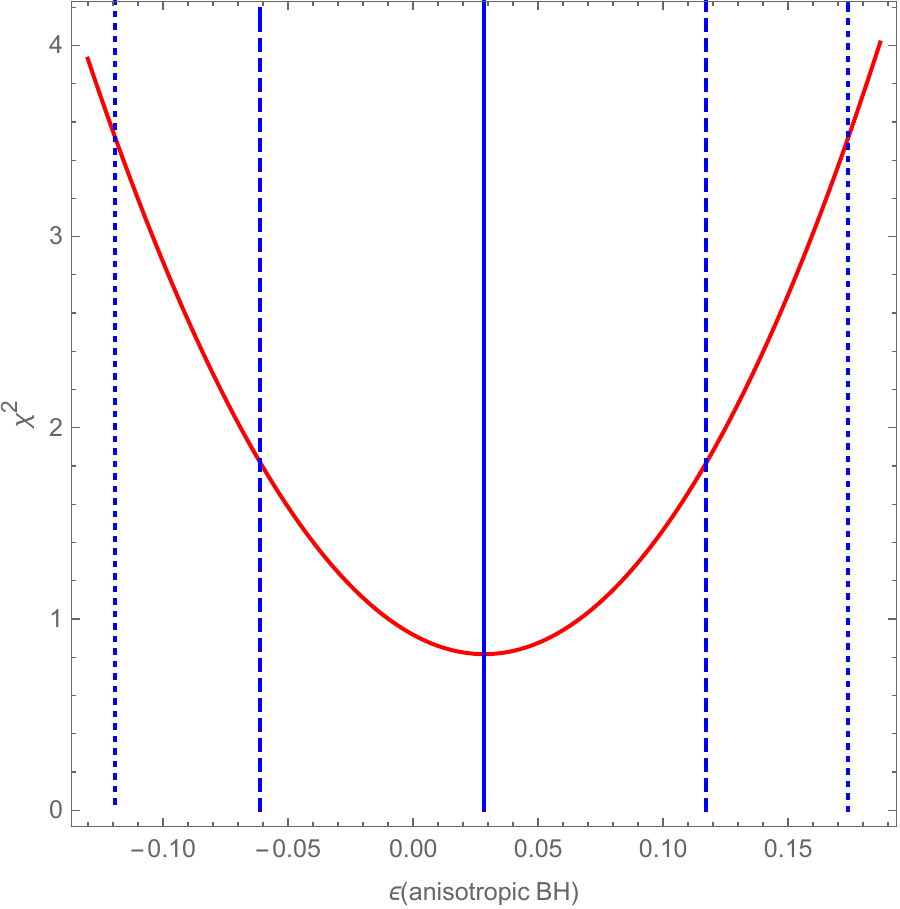}
\includegraphics[width=5cm]{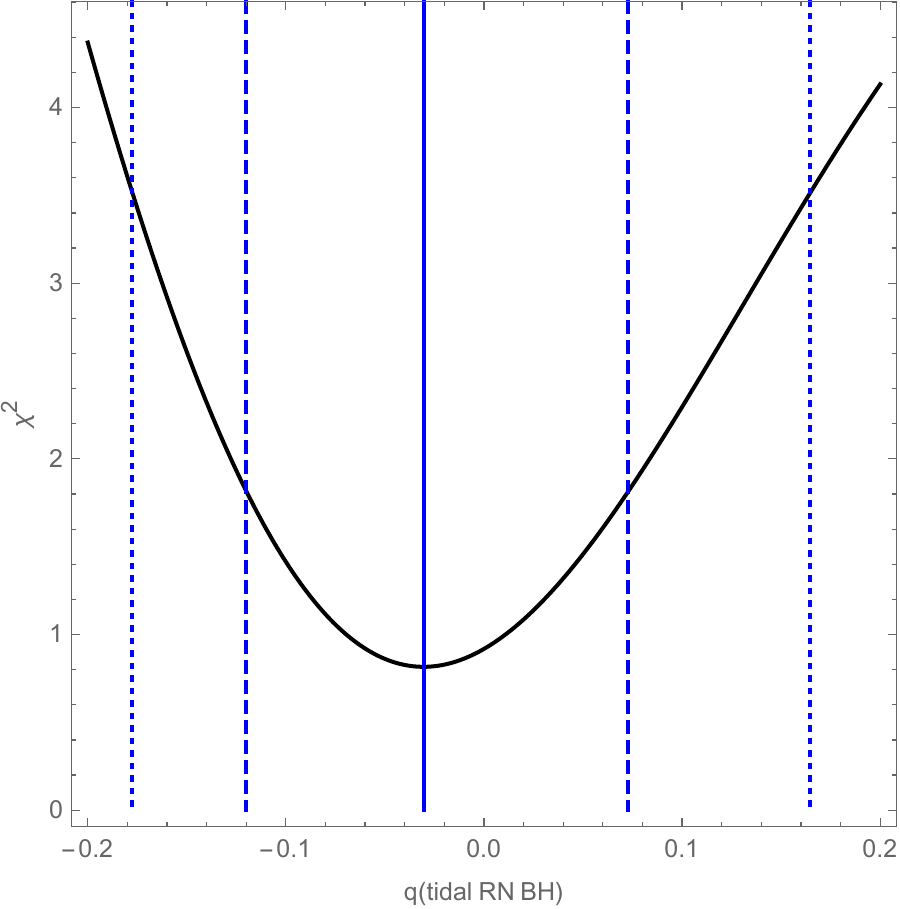}
\caption{The left panel displays    $\chi^2$ for the parameter $\epsilon$ in the anisotropic  black hole. The right panel displays    $\chi^2$ for the parameter $q$       in the  tidal Reissner-Nordstr$\ddot{o}$m  black hole. The  dashed lines denote the $1\sigma$ values. The  dotted lines denote the $2\sigma$ values. The   solid line denote  the best fitted values. }
\label{chi2}
\end{figure}

Therefore, the $\chi^2$ test   on anisotropic metric in the area and isotropic gauge (Eqs.(\ref{area}) and (\ref{iso})) leads to the same parameter ranges. For convenience, we choose the  anisotropic metric in the area  gauge.   The result is summarized in   Figure \ref{chi2}.
For  the  anisotropic black hole,  the observations  show  $\epsilon=0.0285^{+0.0888+0.1456}_{-0.0895-0.1475}$  at $1\sigma$ and $2\sigma$ level of credit confidence. As the physical requirement $\epsilon<0$,  the constrained range $-0.1190<\epsilon<0$  which also satisfies the  mathematical constraint on the near horizon metric is $\epsilon>-1/2$ from the requirement  that  $r_1$ is positive definite \cite{Gregory:2004vt}. For    tidal Reissner-Nordstr$\ddot{o}$m    black hole,   we  obtain   $q=-0.0305^{+0.1034+0.1953}_{-0.0895-0.1470}$  which have the $1\sigma$ and $2\sigma$  regimes of  $q$ with the best-fitted value.  Since our  purpose is to test the braneworld  effect, we constrain the parameter $ -0.1775 <q<0$ in Reissner-Nordstr$\ddot{o}$m    black hole.

  Based on the   $1\sigma$ and $2\sigma$  regimes and  best  fitted values, which are   rowed as $\epsilon_2$, $\epsilon_1$, $q_b$, $q_1$ and $q_2$, we list    the related    observations (including  angular separation, brightness difference, time delay and QNM) in  Tables \ref{tab1}    and \ref{tab2}.  For comparison, these quantities of the Schwarzschild black hole are also listed between the two BHs.  

\begin{figure}[ht]
\centering
\includegraphics[width=5cm]{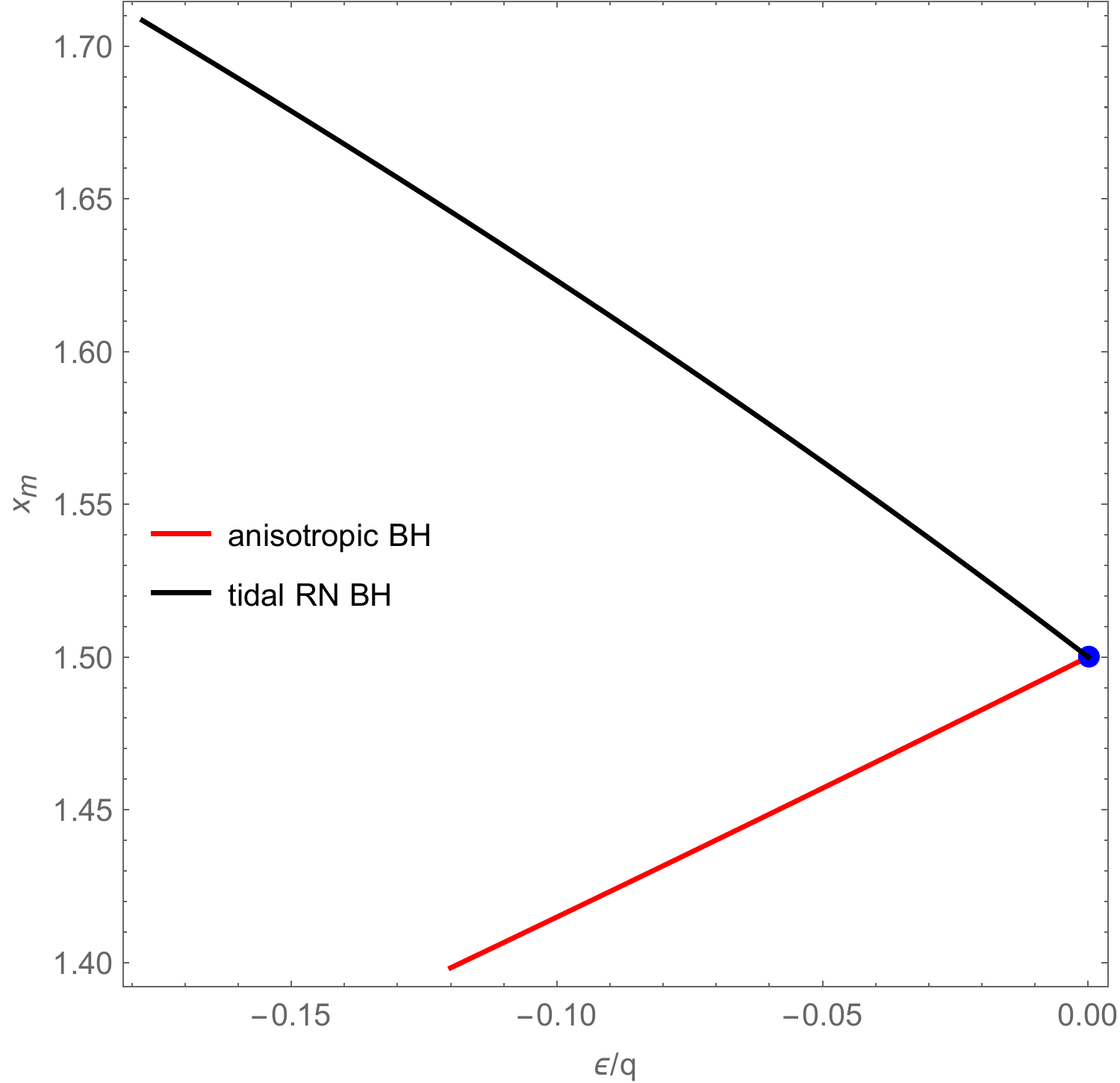}
\includegraphics[width=5cm]{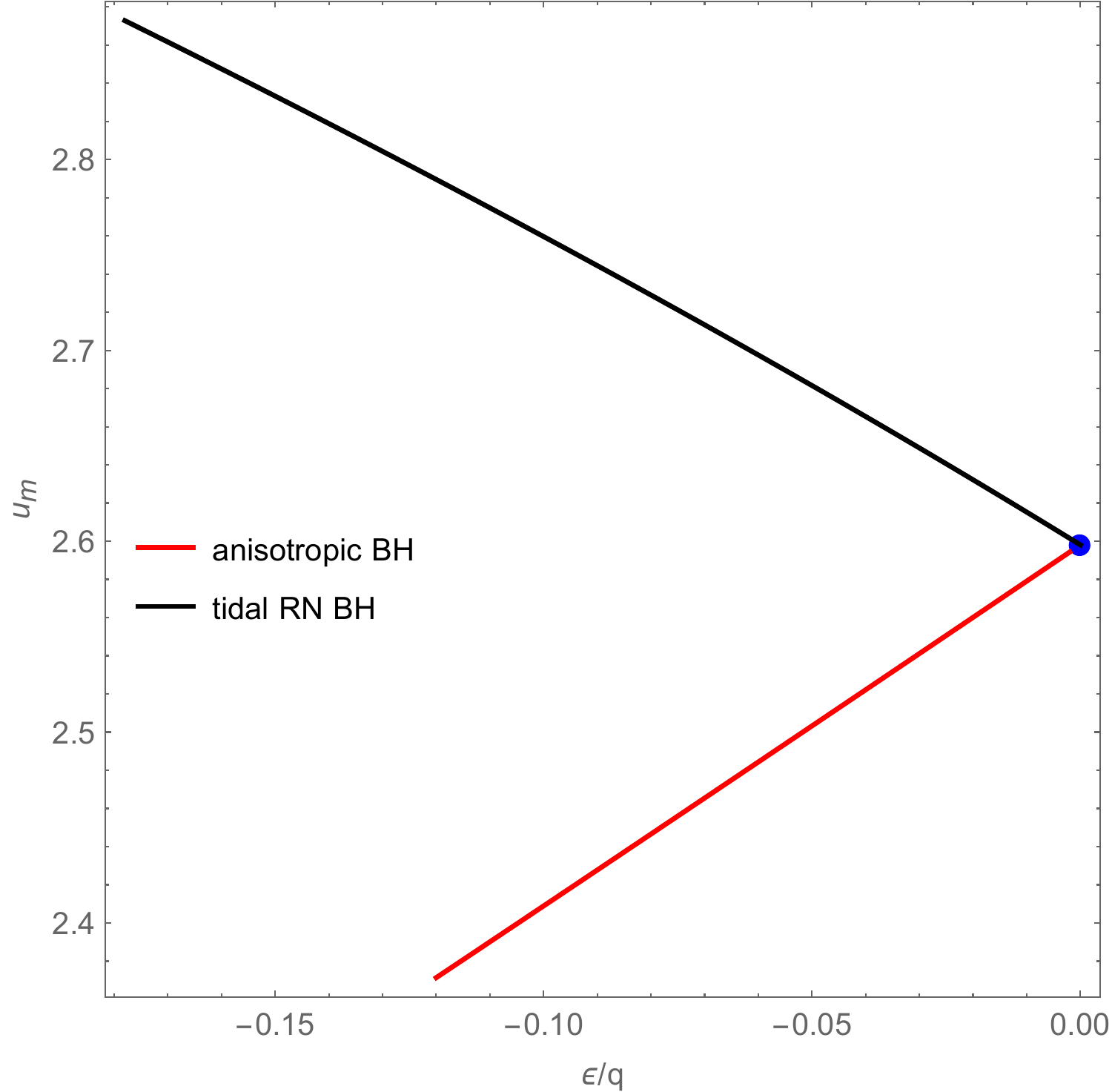}
\includegraphics[width=5cm]{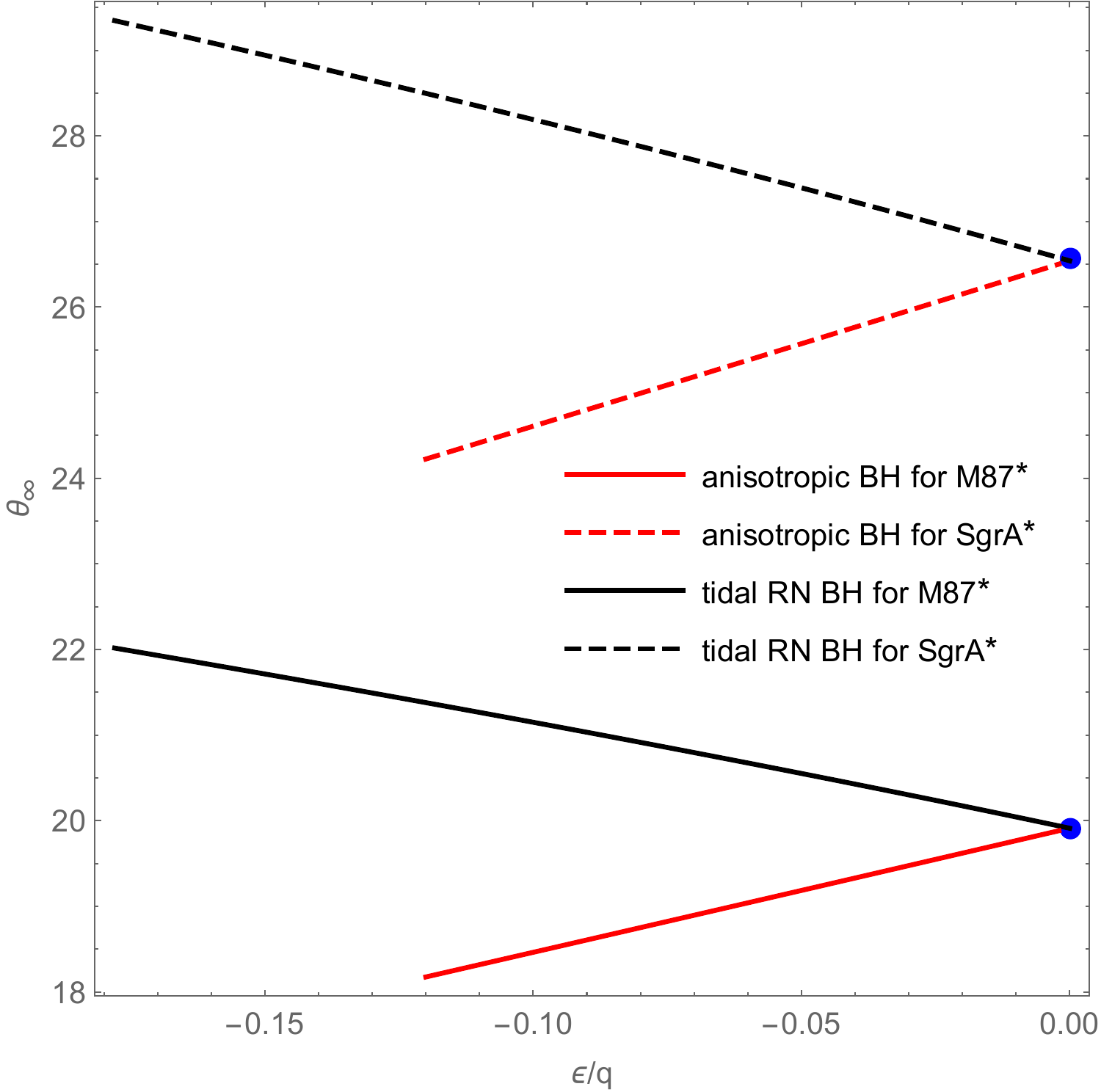}\\
\caption{
The strong deflection lensing parameters $x_m$,$u_m$, and the parameter of observable $\theta_\infty$ vs. $\epsilon$ and  $q$.  The blue dots  denote the Schwarzschild~ case. The $\epsilon$ parameter in the anisotropic  black hole is in the constrained range $-0.1190<\epsilon<0$.  The $q$ parameter in the  tidal Reissner-Nordstr$\ddot{o}$m  black hole is $ -0.1775 <q<0 $.   }
\label{theta}
\end{figure}
\begin{table}[h]
\begin{center}
\small
      \begin{tabular}{c|cc|c|ccc}
   \hline
      \hline
        \multirow{ 2}{*}{Parameters}&\multicolumn{2}{c|}{anisotropic  BH}&  \multirow{ 2}{*}{ SW BH}&\multicolumn{3}{c}{tidal RN BH}\\
    &$  \epsilon_{2}=  -0.1190$ &$\epsilon_{1}= -0.0610 $  &   &$q_{b}= -0.0305$&$q_{1}=  -0.1200   $&$q_{2}=   -0.1775 $\\
     \hline
     $x_m$ &$1.398$&$1.449 $& $1.500 $ &$1.539 $&$  1.645 $&$  1.708 $ \\

\hline
$u_m (R_s)$  &$2.371$&$2.484$&$2.598$ &$2.649 $&$  2.789 $&$  2.872 $\\
$\bar{a}$ &$1.035$ &$1.018$   &$1.000$&$0.9876 $&$  0.9584 $&$  0.9440 $ \\      	
$\bar{b} $ &$-0.4313$ &$-0.4146$&$-0.4002$&$-0.4031 $&$  -0.4132 $&$ -0.4199 $\\

\hline
$\alpha$&$6.498$ &$6.427$ &$6.364$ &$6.296 $&$  6.137 $&$  6.060 $ \\

\hline

  \end{tabular}
    \caption{ The values of parameters $x_m$, $u_m$, $\bar{a}$, $\bar{b} $  and $\alpha$.  The parameters are shown in   $1\sigma$ and $2\sigma$  regimes and  best  fitted values, which are  denoted as  $\epsilon_2$, $\epsilon_1$, $q_b$, $q_1$ and $q_2$. The $u_m$ is scaled by Schwarzschild radius $R_s=2GM_{\bullet}/c^2$.  }
   \label{tab1}
 \end{center}
\end{table}

\begin{figure}[ht]
\centering
\includegraphics[width=5.1cm]{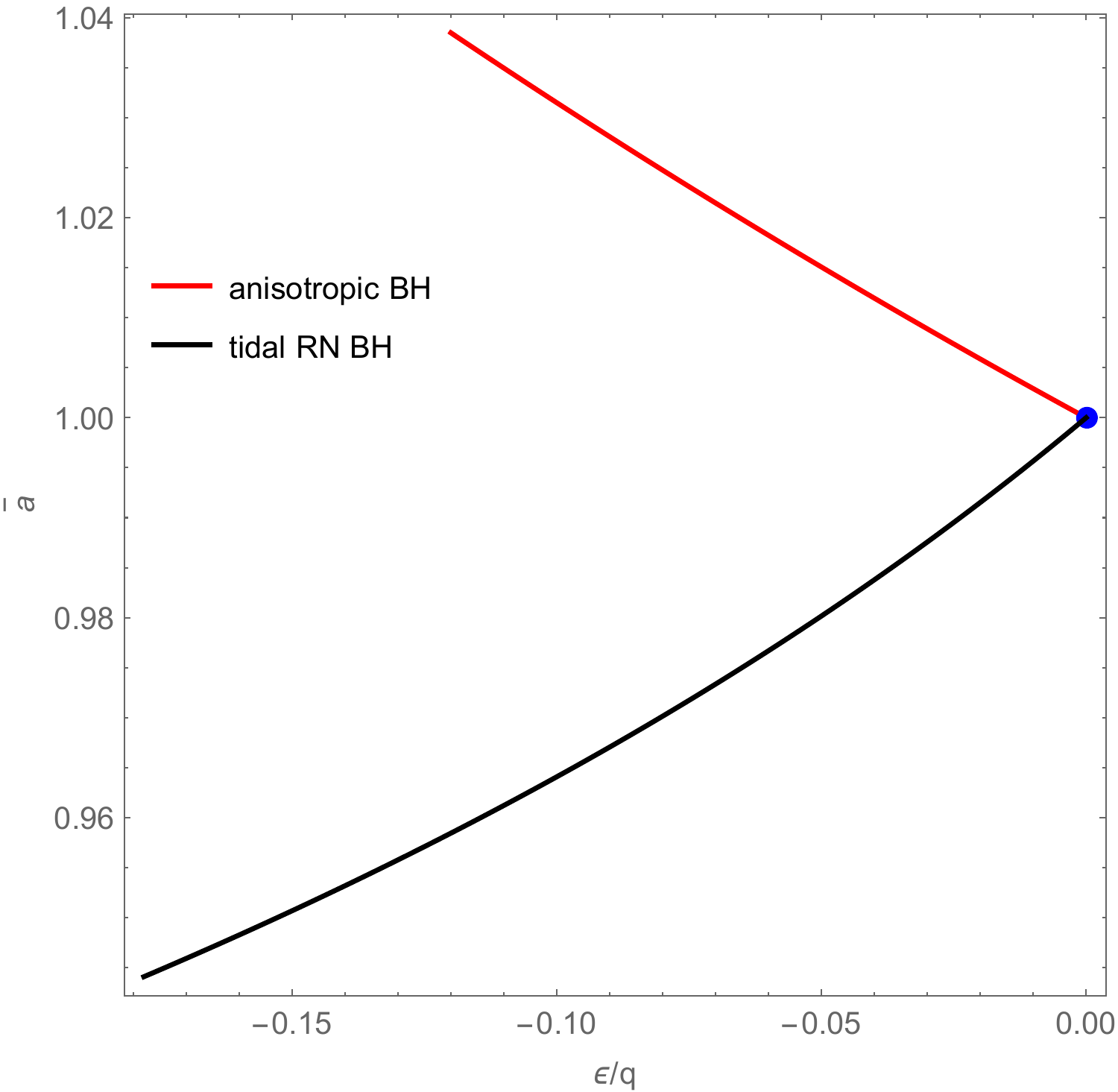}
\includegraphics[width=4.9cm]{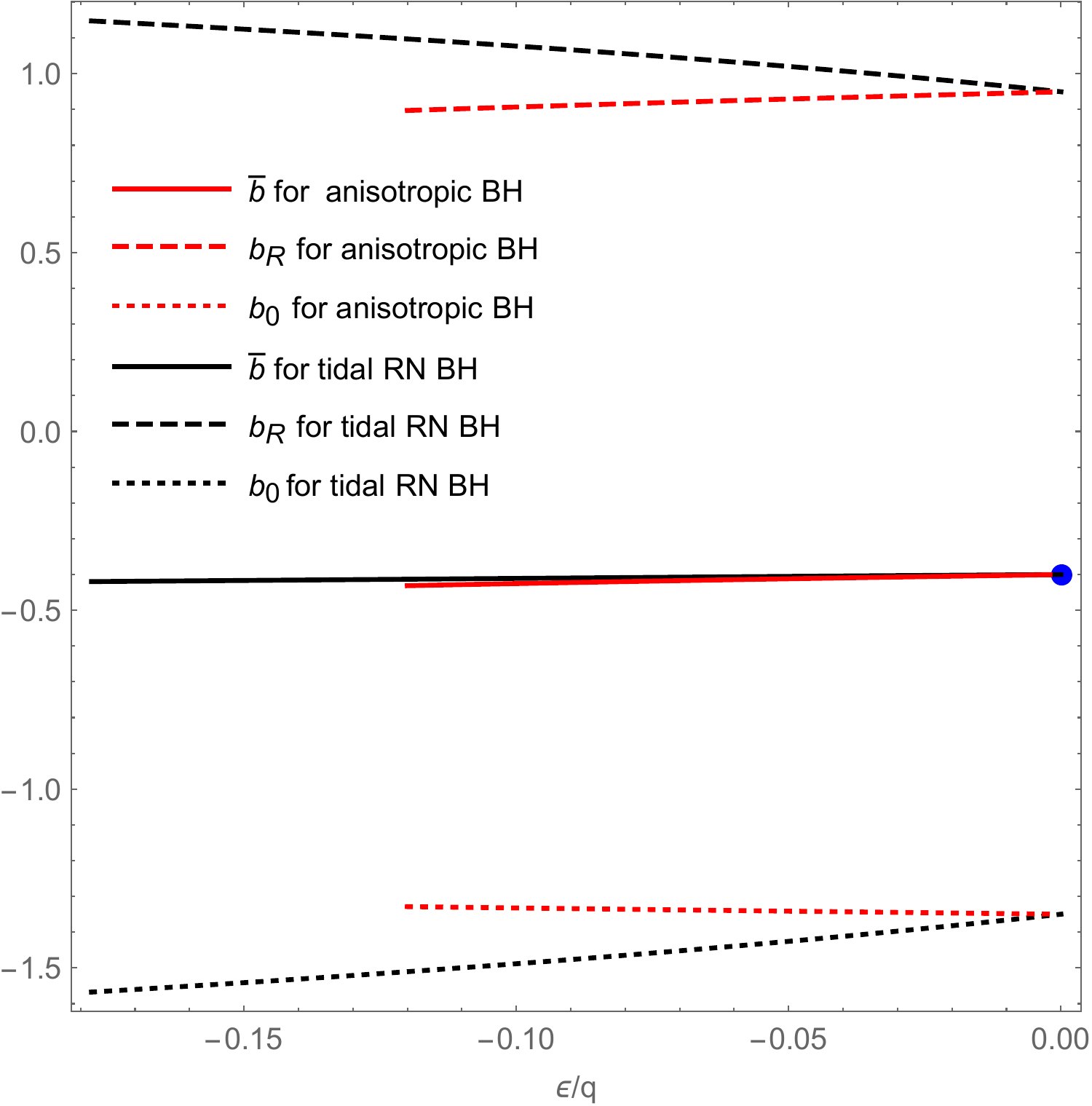}
\includegraphics[width=5cm]{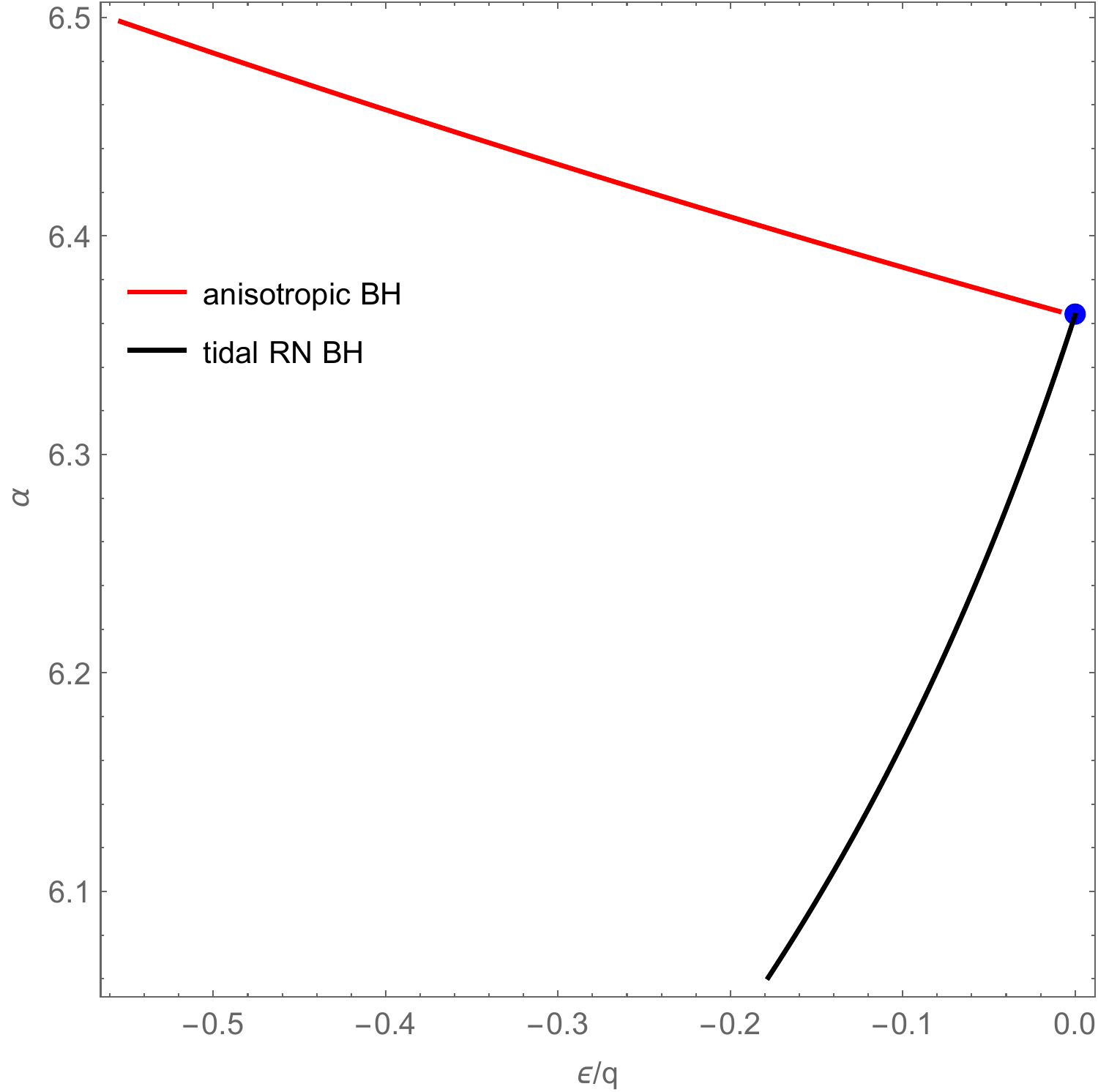}\\
\caption{
The strong deflection lensing coefficients $\bar{a}$,~$\bar{b}$  (including $b_R$, $b_0$  as well) and deflection angle $\alpha$.   The   ranges of parameters are the same as of Figure \ref{theta}.   }
\label{alpha}
\end{figure}
 \subsection{Discussions of the deflection  angle in  theory}

  Firstly,    we list   the deflection  angle related parameter in Table \ref{tab1}. Then,  we plot the  shapes of the parameters  $x_m$, $u_m$ and the observable $\theta_{\infty}$   in   Figure \ref{theta} .  The shapes of them are opposite, with one of which increasing with respect to $\epsilon$ , and the other  decreasing with respect to $q$.   In the case of     tidal Reissner-Nordstr$\ddot{o}$m    black hole,    the $x_m$, $u_m$ and $\theta_{\infty}$ parameters  are always smaller than the one  of  the Schwarzschild black hole with the same mass. While in  the anisotropic black hole,  $x_m$, $u_m$ and $\theta_{\infty}$ are smaller than the ones in Schwarzschild~ black hole.   The Schwarzchild~ black hole  connects  the two  braneworld black holes which denotes that we could not  distinguish the two braneworld black holes from the Schwarzchild~ black hole.
As  $u_m$ is determined  by the non-linear relation  (Eq.(\ref{um})) with $x_m$,      the  slope of $u_m$ for  the anisotropic  black hole  is smaller  than that  of the   tidal Reissner-Nordstr$\ddot{o}$m    black hole.

In Figure  \ref{alpha},
 the parameters $\bar{a}$ and $\bar{b}$ play a prominent role in measuring the angular difference from the outmost image and the adherent point related to the sequence of subsequent images.
 Roughly speaking, a bigger $u_m$  implies a smaller $\alpha$.
Then the  shape of $\bar{a}$ determines the shape of $\alpha$ which  represents the divergence part.  And the first  term of  $\alpha$ is more important than the other ones.
   Furthermore,  corresponding to Eqs.(\ref{a}) and (\ref{b}),    the divergent parts $\bar{a}$ and $b_0$  have  a decreasing tendency, while the regular part  $b_R$  contributes to the increasing  tendency.
The parameter  $\bar{b}$, which presents the main part of the regular part, is one order lower than  $\alpha$, and then its non-monotonic  value does not affect $\alpha$. But, the shape of $\alpha$ parameter is  not smooth as that of $\bar{a}$.



\begin{table}[h]
\small
\begin{center}
      \begin{tabular}{c|c|cc|c|ccc}
   \hline
      \hline
        \multirow{ 2}{*}{Parameters}&  \multirow{ 2}{*}{BH}&\multicolumn{2}{c|}{anisotropic  BH}&  \multirow{ 2}{*}{ SW BH}&\multicolumn{3}{c}{tidal RN BH}\\
    &&$  \epsilon_{2}=  -0.1190$ &$\epsilon_{1}= -0.0610 $ &   &$q_{b}= -0.0305$&$q_{1}=  -0.1200   $&$q_{2}=  -0.1775 $\\
\hline
     \multirow{ 2}{*}{$\theta_\infty ~(\mu as)    $}& M87*&$18.17$&$19.04$ &$19.91  $ &$20.30 $&$  21.38 $&$  22.02$\\
     &Sgr $A^*$&$24.22$&$25.38$& $26.54  $ &$27.06 $&$  28.50 $&$  29.32 $\\
  \multirow{ 2}{*}{$s~(nas) $}    &  M87*&$28.28$&$26.48$&$24.92  $  & $23.29 $&$  19.75 $&$  18.6 $\\
  &Sgr  $A^*$&$37.70$&$35.30$&$33.22$  &$31.04 $&$  26.33 $&$  24.21 $ \\
$r~(mag)$ &$-$&$6.569$ &$6.700$& $6.822$   &$6.908 $&$  7.118 $&$ 7.226 $ \\

   \hline
$\Omega_m$&$-$&$0.4217$&$0.4025$&$ 0.3849 $  &  $0.3775 $&$  0.3585 $&$  0.3481 $\\
$\lambda$&$-$&$0.1263$&$0.1219$&$0.1176$   & $0.1161 $&$  0.1120 $&$  0.1096 $\\
\hline
$\Delta  T_{2,1}    $&$-$&$15.16$&$15.86$&$16.57$&$16.88 $&$  17.75 $&$  18.26 $\\
$\Delta T_{3,2}     $&$-$&$14.91$ &$15.62$&$16.33$&$16.65 $&$  17.54 $&$  18.06 $\\
$\Delta T_{4,3}     $&$-$&$14.90$&$15.61$&$16.32$&$16.64 $&$  17.53 $&$  18.05 $\\
\hline
$\Delta  T^1_{2,1}(10^{-3})   $&$-$&$261.4$&$  248.7$& $248.7 $ &  $240.2 $&$  221.2 $&$  212.4 $\\
$\Delta T^1_{3,2}(10^{-5})    $&$-$&$1269$ &$1164$  & $1075$ &$998.0 $&$  834.0 $&$  762.0 $\\
$\Delta T^1_{4,3}(10^{-7})    $&$-$&$6163  $ &$5322$&$4645$&  $ 4144 $&$  3146 $&$  2734 $\\
\hline

  \end{tabular}
    \caption{ The values of observables $\theta_{\infty}$, $s$, $r$,  the QNM and the time delay    parameters  based on the   $1\sigma$ and $2\sigma$  regimes and  best  fitted values, which are  denoted as  $\epsilon_2$, $\epsilon_1$, $q_b$, $q_1$ and $q_2$.  The $\theta_{\infty}$ and $s$   are respectively in  units of micro-arcsecond ($\mu$as) and nano-arcsecond (nas).
 The time delay   parameters $\Delta T_{n,m}$ and $\Delta T_{n,m}^{\ 1}$ are scaled by $2GM_{\bullet}/c^{3}\approx42.45s$.  }   \label{tab2}
 \end{center}
\end{table}
\subsection{The observations}
Besides $\theta_{\infty}$, there are other relations which could translate the  parameters  $\bar{a}$ and $\bar{b}$ to the observables, e.g. $s$, $r$, the time delays and the QNMs. We list  all observables in Table \ref{tab2} which show the same tendencies.

\subsubsection{The $s$ and $r$   parameters}
\begin{figure}[ht]

\centering
\includegraphics[width=5cm]{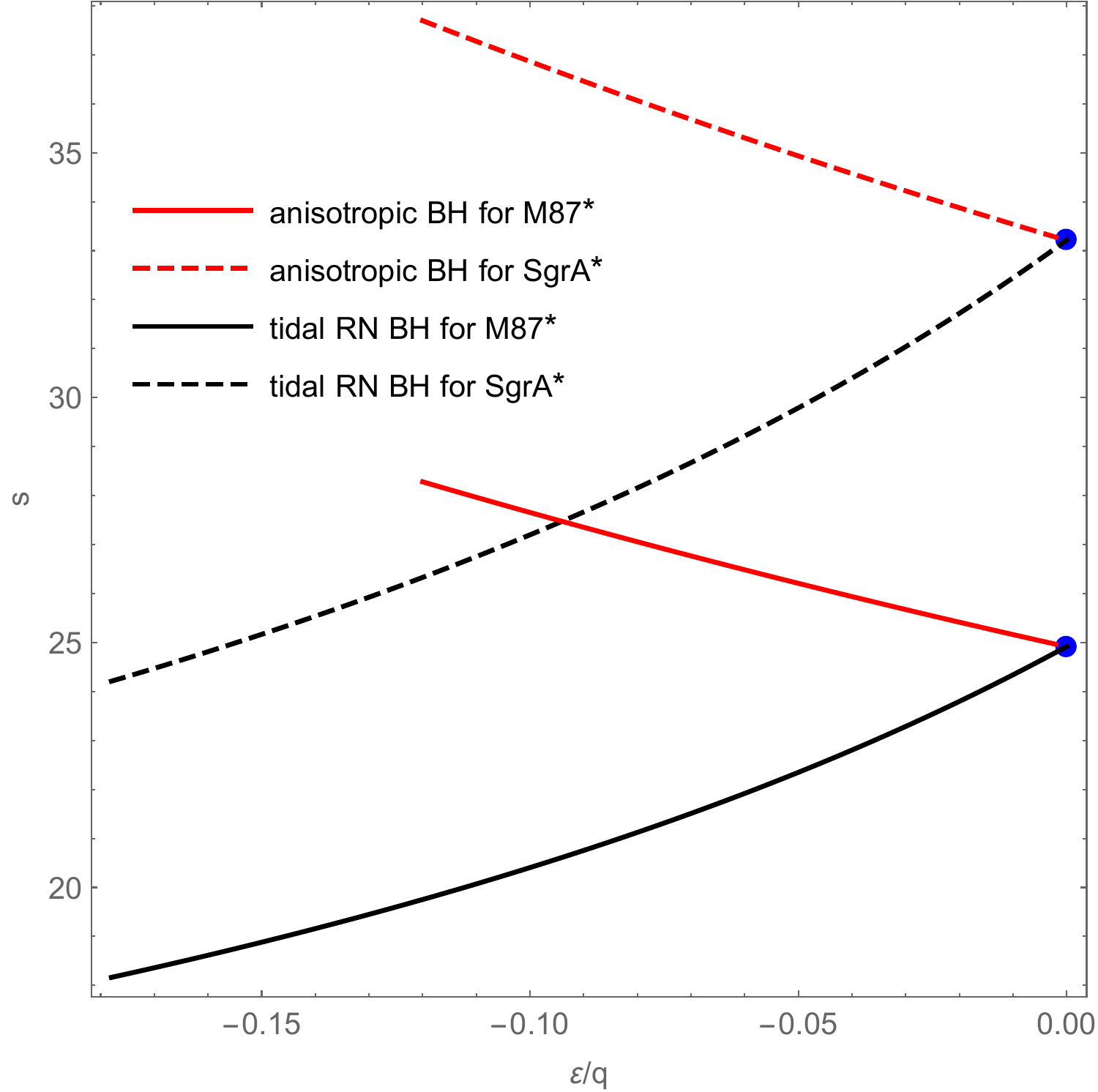}
\includegraphics[width=5cm]{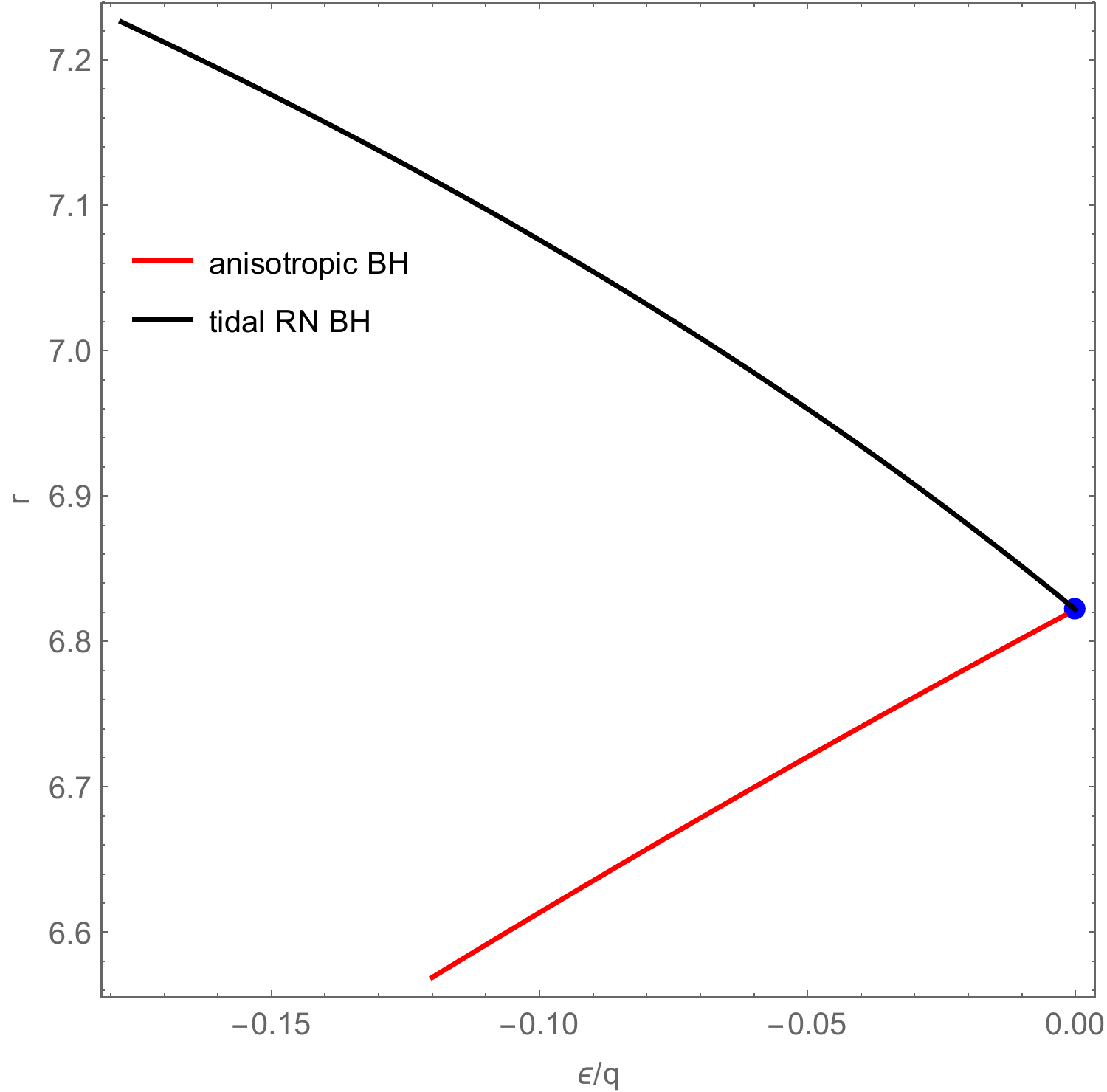}\\
\caption{The estimated observables ~$s$,~$r$ as functions of $\epsilon$ (or $q$) for Anistropic Black hole (or     tidal Reissner-Nordstr$\ddot{o}$m    Black Hole).  The   ranges of parameters are the same as in    Figure.\ref{theta}. }
\label{sr}
\end{figure}

The observable $s$ is the angular separation between the outermost image ($n=1$) and other packed  $n=2,3,....\infty$, and $r$ is the magnitude difference between the outermost image and the packed images,
  \begin{eqnarray}
  \label{s}
&&s=\theta_1-\theta_{\infty}=\theta_{\infty}\exp{(\frac{\bar{b}}{\bar{a}}-\frac{2\pi}{\bar{a}})},\\
&&r=2.5\log_{10} (\frac{\mu_1}{\sum_{n=2}^{\infty}\mu_n})=2.5\log_{10} [\exp{(\frac{2\pi}{\bar{a}})}].
\end{eqnarray}
  We have plotted the observable parameters $s$ and $r$ in Figure \ref{sr} .
From  Figure \ref{sr}, the angular separation $s$ increases while angular position ($\theta_{\infty}$) and flux magnitude ($\mu$) decrease with respect to $\epsilon$ and $q$. The parameter  $s$ is much smaller than $\theta_{\infty}$($1/1000$).
As  shown in Eq.(\ref{s}), $\theta_1$  is approaching  $\theta_{\infty}$.  That means the $\theta_{\infty}$ parameter could present the main effect of  angular  separation.
 As Ref. \cite{Bozza:2008ev} shows, the small angles lens equation (Eq.(\ref{smallangle})) brings an error of $\theta_{\infty}$  about $GM/3D_{LS}$ which $0.07\mu as$   for    Sgr. A* and $0.05\mu as$   for M87*.  Then by comparing  the  $\theta_{\infty}$s listed in Table  \ref{tab2} , this  form of lens equation leads to about $5\%-6\%$  systematic error of $\theta_{\infty}$ in $1\sigma$ range (about  $3\%$  systematic error in $2\sigma$ range). This error  should be considered in future constraints.  The parameter $s$ in anisotropic black hole is nearly linear, while  in    tidal Reissner-Nordstr$\ddot{o}$m    black hole is non-linear.  Our results of $s$ are  consistent  with Ref.{\cite{Bin-Nun:2009hct}} where $\theta_2-\theta_1=0.03 \mu as$ for tidal RN BH. But it is hard to observe $s$ and $r$ because the magnitude ratio is proportional to the magnitude.
 The $s$ is increasing with respect   to $q$,  decreasing with respect   to $\epsilon$ ,while the $r$ is the opposite tendency.   If we try to distinguish them via observation data, the accuracy of the measured separation between the first image and the surplus fringes needs to be less than $1\mu as$, and the photometric uncertainty has to be better than $0.1$ mag.

\begin{figure}[htbp]
\centering\includegraphics[width=6cm]{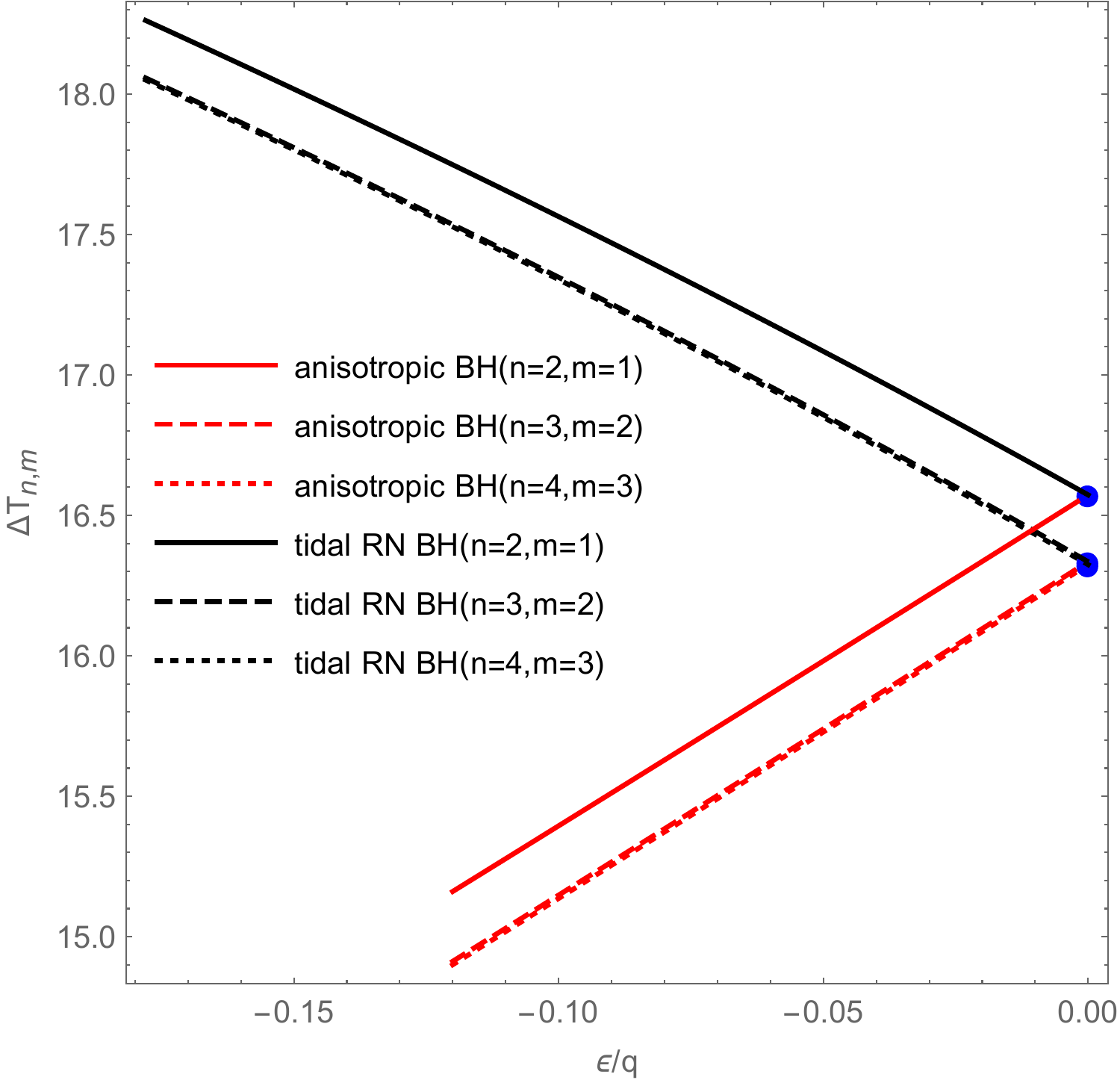}
\includegraphics[width=6cm]{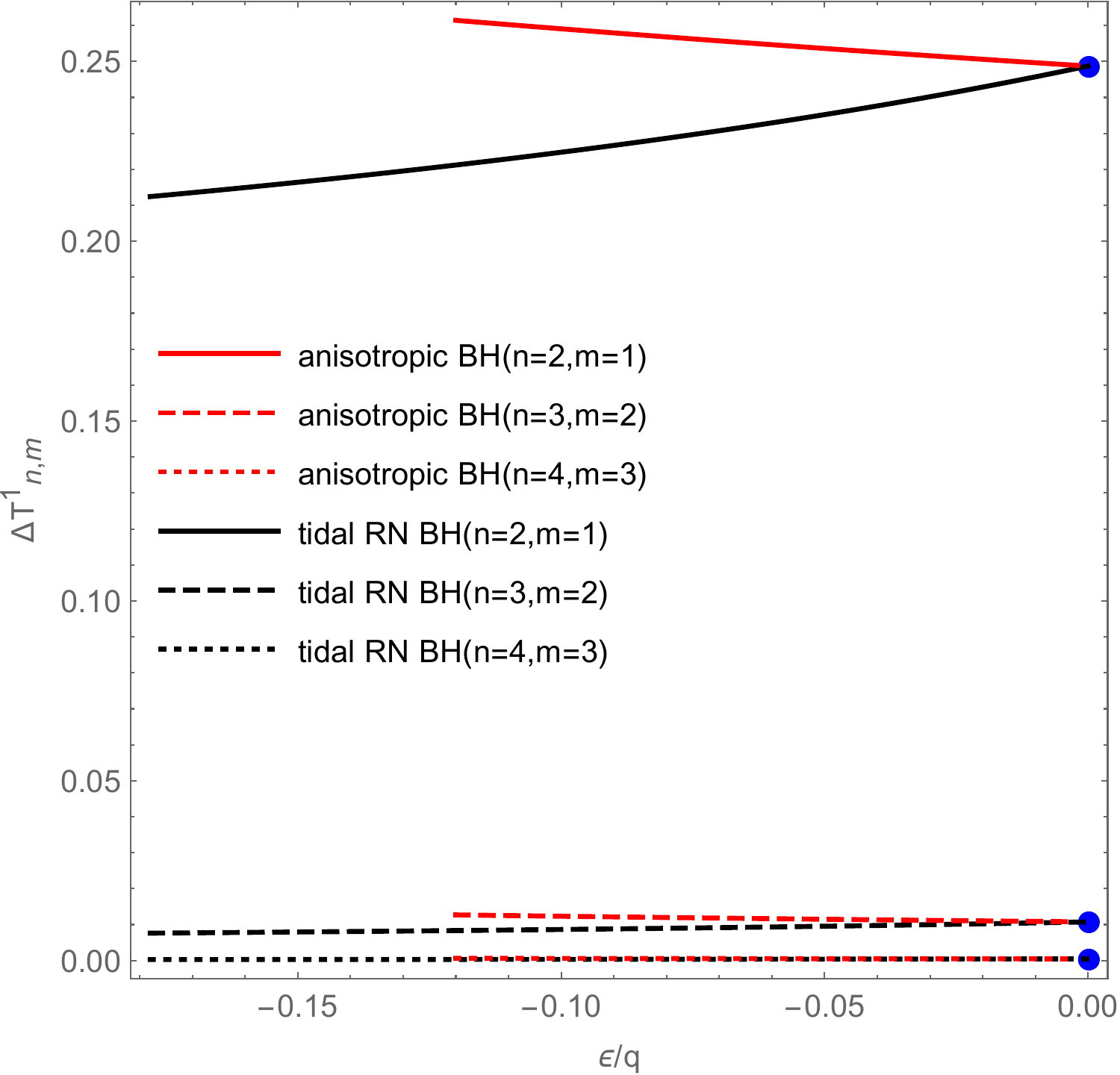}\\

\caption{  The estimated observables $\Delta T_{n,m}$ and $\Delta T_{n,m}^{\ 1}$  as functions of $\epsilon$ (or $q$) for anisotropic Black hole (or     tidal Reissner-Nordstr$\ddot{o}$m    Black Hole).  The   ranges of  obervables are the same as in   Figure \ref{theta}.   $\Delta T_{n,m}$ and $\Delta T_{n,m}^{\ 1}$ are both represented in the unit of $2GM_{\bullet}/c^{3}\approx42.45$ s. }
\label{td}
\end{figure}

\subsubsection{ The  time delay }
The time delays between relativistic images are  distinguished     as well.
Bozza  and Mancini obtain  different time delays among  relativistic images due to  gravitational lensing  by a general static spherically  symmetry spacetime  \cite{Bozza:2003cp}.     The time delay can attribute to different paths  followed by the photons when they cross around the black hole.
Differences in the deflection angle are immediately displayed on the relativistic images\cite{Bozza:2003cp}.  If the mass and distance of the lens ( $D_{os}$) are known, then any set of relativistic images could probe the type of   black hole.
$\Delta   T_{n,m}$ is the total time delay between the $m$-loop image and the $n$-loop image
\begin{eqnarray}
\Delta T_{n,m}=\Delta T^0_{n,m}+\Delta T^1_{n,m},
\end{eqnarray}
where the  leading term of time delay is
\begin{eqnarray}
\Delta T^0_{n,m}=2\pi(n-m)u_m,
\end{eqnarray}
 while  its much smaller correction is
  \begin{eqnarray}
  \Delta T^1_{n,m}=2\sqrt{B_m/A_m}\sqrt{u_m/\hat{c}}\exp{(\bar{b}/(2\bar{a}))}[\exp{(-m\pi/\bar{a})}-\exp{(n\pi/\bar{a})}].
  \end{eqnarray}
 The unit of time delay is  $2GM_{\bullet}/c^{3}  \sim 42.45$ s.
The dominant term in the time delay  is not a new independent factor of the black hole,  but the second term is.
We consider three    cases  ( ($n=2,m=1$), ($n=3,m=2$), ($n=4,m=3$)) satisfying $n-m=1$ in which we consider two nearby loops  and set the  same $T^0_{n,m}$.   As  shown in    Figure \ref{td}, the time delay in the two nearby loops is  from  $\Delta T^1_{n,m}$ which decreases to $0$ as fast  as $n$ increases.
The same tendency occurs in the leading term $T_{n,m}$ when $n-m=1$, which is consistent  with small $s$. The phenomenon shows there is no significant difference between  the outmost image and the stacked images. Our present   observational facilities do not reach the required resolution yet. For galactic black hole, the required resolution is of the order of $0.01$ micro-arcsecs.  As  the unit is $2GM_{\bullet}/c^{3}  \sim 42.45$,  the detection of $\Delta T_{2,1}$ needs to have an accuracy better than the level of $\sim 2\times 10^{-2}$ .  After multiplying     the units, it corresponds to the level of about $1s$ for Sgr A*.

\subsubsection{The quasi normal modes (QNMs)}
\begin{figure}[htbp]
\centering
\includegraphics[width=6cm]{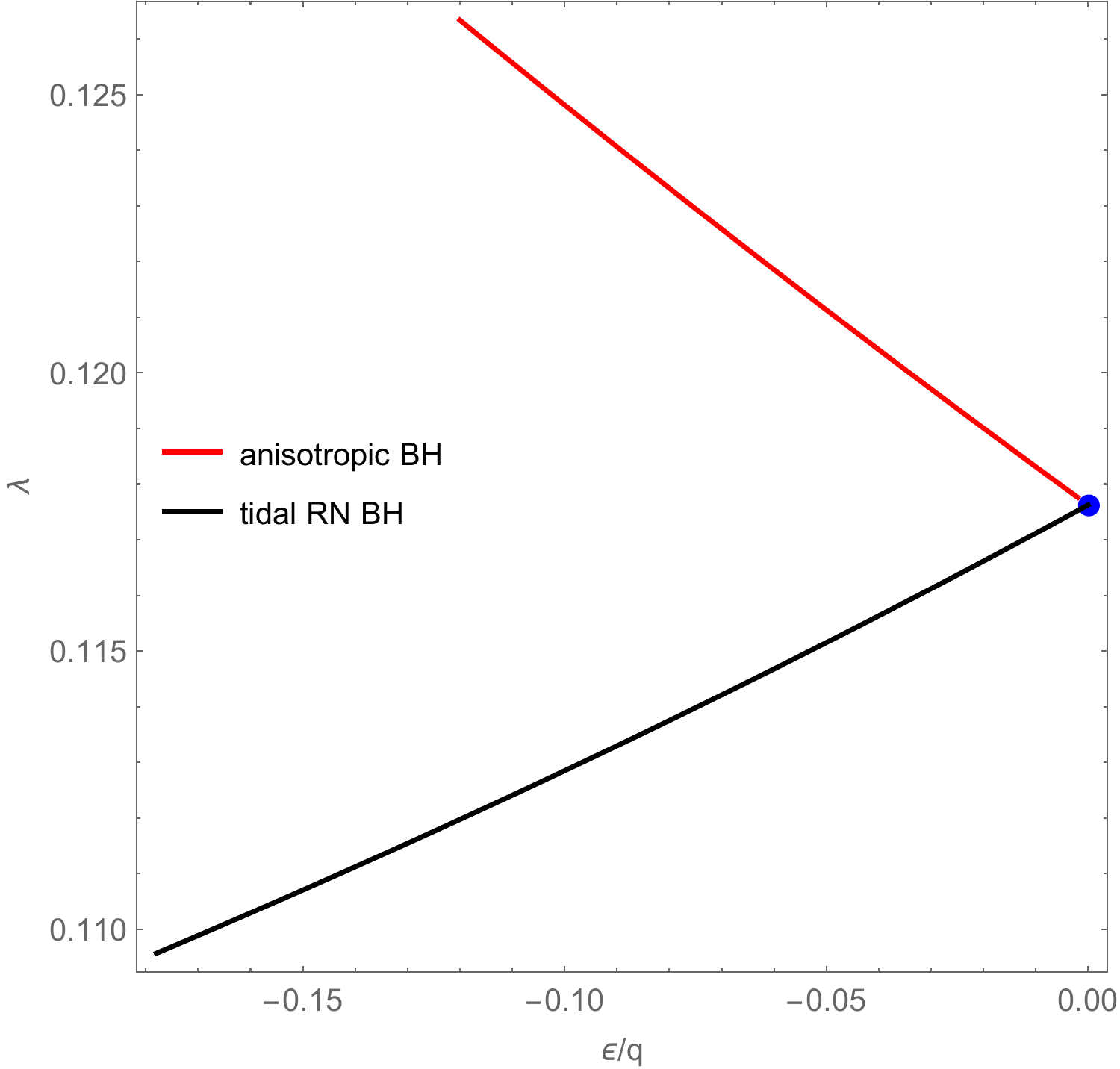}
\includegraphics[width=6cm] {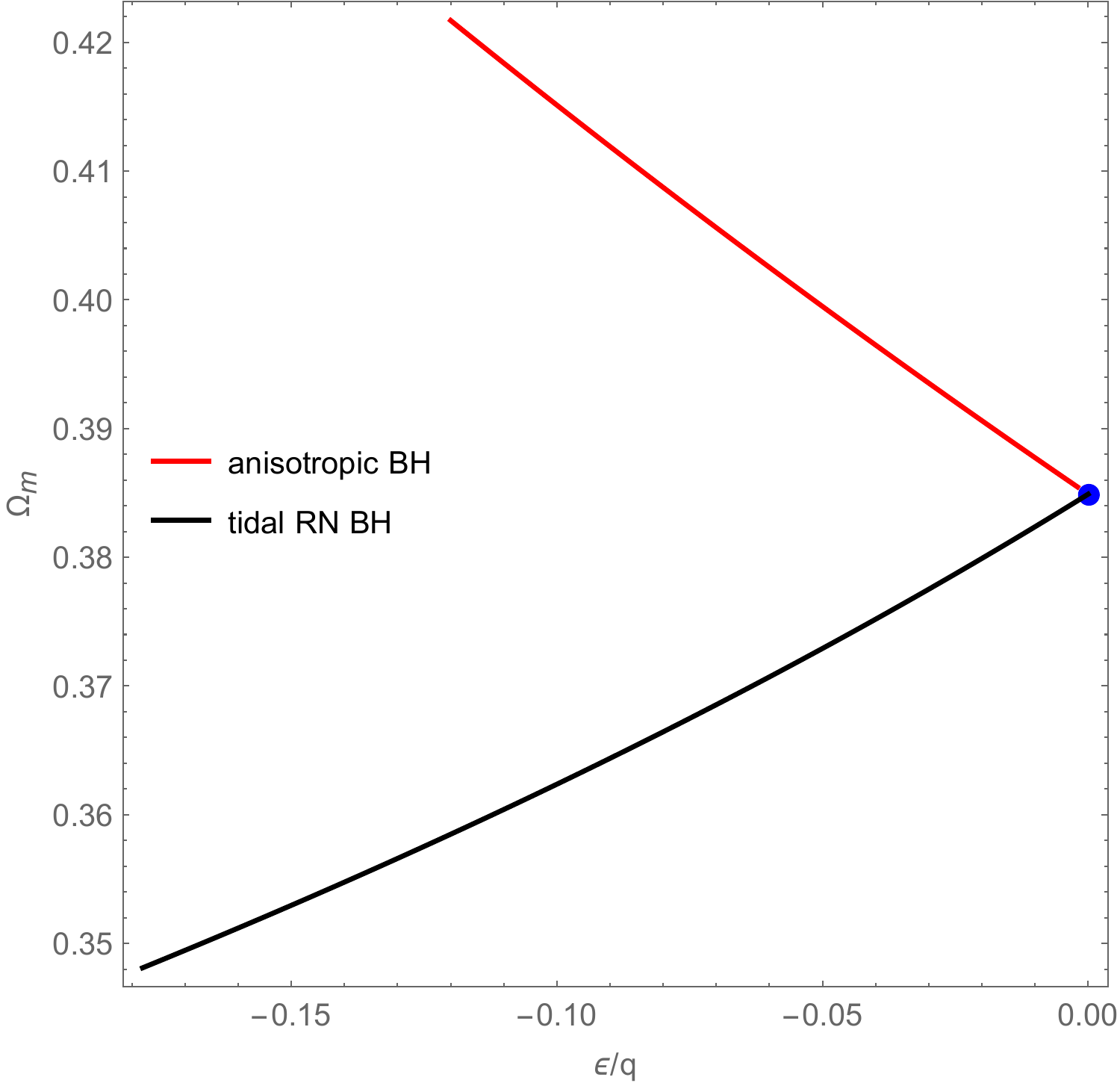}
\caption{  The estimated observables $\lambda$  and $\Omega_{m}$  as functions of $\epsilon$ (or $q$) for anisotropic black hole (or     tidal Reissner-Nordstr$\ddot{o}$m    Black Hole).   The   ranges of parameters are the same as in   Figure  \ref{theta}.  }
\label{qnm}
\end{figure}

The strong lensing is useful to explain the  characteristic  modes  of  black hole as well.
 In Refs.  \cite{Cardoso:2008bp,Dolan:2009nk,Stefanov:2010xz,Bohra:2023vls},  the quasi normal modes (QNMs)  and the strong  lensing   are found to connect with each other.    The QNM describes the decay rates of perturbations around a black hole.  It is expected to detect these perturbations  in  further observations.
 At eikonal limit, the  real and imaginary parts of the QNMs of any spherically symmetric, asymptotically flat  spacetime are given by  (multiples of)  the frequency ($\Omega_m$) and instability timescale of the unstable circular null geodesics.
 \begin{eqnarray}
\omega_{QNM}=\Omega_ml-i(n+1/2)|\lambda|,
\end{eqnarray}
where $l$  and $n$ are constants and
 \begin{eqnarray}
&&\lambda=\frac{1}{u_m\bar{a}},\\
&&\Omega_m=1/u_m.
\end{eqnarray}
 The real  part of the complex QNM frequencies is determined by the angular velocity at the unstable null geodesic, and the imaginary part of the QNM is related to the instability  timescale of the orbit which is called  the Lyapunov exponent. The  Lyapunov exponent  is in  turn reflected  in the associated QNMs in the geometrical optical approximation.
We plot the values of QNMs in Figure \ref{qnm}.  The tendencies of $\lambda$ and $\Omega_m$  are similar.   For the  Schwarzchild~ model, it is the  dot  at $q=0$, while for the two braneworld black holes, it  decreases with respect to $\epsilon$ and $q$.
  Our constrained    Lyapunov   exponents for M87* and SgrA*  are positive.
A positive Lyapunov exponent indicates a divergence between nearby trajectories, i.e., a high sensitivity to initial conditions.

\subsection{A short summary}

To derive  the deflection angle, we need three parameters  ($u_m$, $\bar{a}$  and  $\bar{b}$).
The observations $\theta_{\infty}$, $T_{n,m}$ and $\Omega_{m}$ are  all related to $u_m$.    The parameters $\lambda$  and $r$ are related to $\bar{a}$. The parameters $s$  and $\Delta  T^1_{n,m}$ are related to  $\bar{b}$.  The higher order effects, such as $s$,  $r$, $\Delta T^1_{2,1}$ will distinguish   the two black holes.  We also show the total effect of $\alpha$.   The non-linear relation between $u$ and $x_m$ makes the  non-smoothness between  the  braneworld  models.


\section{Conclusion }\label{conclusion}

 In this work,  based on  the  EHT data ($\theta_\infty$), we first use the $\chi^2$ test  to   estimate    the  range of parameters of braneworld black holes. Then,  the  $1\sigma$ and  $2\sigma$ regimes  of the model parameters are  $\epsilon=0.0285^{+0.0888+0.1456}_{-0.0895-0.1475}$ for the anisotropic black hole, and $q=-0.0305^{+0.1034+0.1953}_{-0.0895-0.1470}$ for the    tidal Reissner-Nordstr$\ddot{o}$m    black hole.   Based on  the fitted data and physical requirement, we calculate the photon deflection, the angular separation and time delay of different relativistic images of the  anisotropic black hole  and the tidal RN black hole in  the  ranges $-0.1190<\epsilon<0$ and  $-0.1775 <q<0$.  The braneworld model is consistent with the observation which shows the braneworld black holes possess  richer structure than ordinary black holes.   And following  the fitted data, we calculate the photon deflection angle, the angular separation, time delay values and QNM values of different relativistic images of the anisotropic black hole and the     tidal Reissner-Nordstr$\ddot{o}$m    black hole.  Our results shed light for probing  extra dimensions.

\acknowledgments
YZ is supported by National Natural Science Foundation of China under Grant No.12275037   and 12275106.
DW is supported by the NSFC under Grants No. 12205032, CQ RLSBJ under grant cx2021044,
and the Talent Introduction Program of Chongqing University of Posts and Telecommunications under grant No. E012A2020248. HZ  is supported by  National Natural Science Foundation of China Grants Nos.12275106 and 12235019.


\end{document}